# Chrysalis Cipher Suite

Ian Malloy

Dennis Hollenbeck

Abstract

The formal verification of architectural strength in terms of computational complexity is achieved through reduction of the Non-Commutative Grothendieck problem in the form of a quadratic lattice. This multivariate form relies on equivalences derived from a k-clique problem within a multigraph. The proposed scheme reduces the k-clique problem as an input function, resulting in the generation of a quadratic used as parameters for the lattice. By Grothendieck's inequality, the satisfiability of lattice constraints in terms of NP-Hard and NP-Complete bounds is provably congruent to a closest vector problem in the lattice. The base vectors of the resulting lattice are treated as a holomorphic vector bundle. From the resulting bilinear matrices, the tight-hardness reduction of the closest vector problem as the shortest vector problem is introduced within the system. The derivation of the closest vector problem requires that the lattice is necessarily generated by a $\binom{0}{1}$-Matrix expressed as a quadratic. This vector bundle is denoted as the unit ball with congruent topology to the Riemann sphere, symbolized as $\mathcal{O}$. For the Grothendieck constraints, the relative vector norms necessarily result in satisfaction of NP-Hard requirements for shortest vector problems in the lattice.

The manifold $(M)$ shall be defined as a closed and open set, using the principles of holomorphic vector bundles to introduce a Hermitian metric within the complex vector space. The tangent bundle is by definition a disjoint union of the tangent spaces of the manifold, of which the maximal expression is the complex vector space. The image$(f^{\star})$, and preimage $(f_{\star})$ derived from the same function, is based upon a substitution in the principle holomorphic function. The differentiable manifold allows for multivariate analysis which lends itself towards the principle bundle of frames relative to the tangent bundle.

The security reduction proposed is the NP-Hard Non-Commutative Grothendieck problem which states that "…for any $\varepsilon > 0$ it is NP-Hard to approximate the non-commutative Grothendieck problem to within a factor of $\frac{1}{2} + \varepsilon$" (Briet, Regev, & Saket, 2015). Hermitian matrices may be derived as Hermitian matrices of norm 1 with a loss factor of $\sqrt{2}$ in the approximation but still allow for proof of the non-commutative Grothendieck inequality and has been demonstrated algorithmically (Naor, Regev, & Vidick, 2012).

Applications to a dense regularity for algorithmic non-commutation result in recasting the problem as a semidefinite program (Naor, Regev, & Vidick, 2012). A linear map in *n-dimensional* $\mathbb{C}$, or $\mathbb{C}^n$ to any Banach space reframes the problem computing the norm of the linear map for any $\varepsilon' < 0$ as positive integer $(n)$, such that it is NP-Hard to approximate the norm of the explicitly given linear operator $\mathcal{F}: L_2 \rightarrow L_2(X_n)$ to within a factor greater than $\frac{\tau}{\eta} + \varepsilon'$ (Briet Regev, Saket).

From Briet, et. al.:

"Let $(X_n)_{n \in \mathbb{N}}$ be a family of finite-dimensional Banach spaces, and $\eta$ and $\tau$ be positive numbers such that $\eta > \tau$. Suppose that for each positive integer $(n)$ there exists a linear operator $f: \mathbb{C}^n \rightarrow X_n$ with the following properties:

- For any vector $a \in \mathbb{C}^n$, we have $\|f(a)\|_{X_n} \leq \|a\|_{\ell_2}$
- For each standard basis vector $e_i$, we have $\|f(e_i\|_{X_n} \geq \eta$

Then, for any $\varepsilon' > 0$ there exists a positive integer $n$ such that it is NP-Hard to approximate the norm of an explicitly linear operator $\mathcal{F}: L_2 \rightarrow L_1(X_n)$ to within a factor greater than$\left(\frac{\tau}{\eta}\right) + \varepsilon'$." (Briet, Regev, & Saket, 2015).

The complex case of these properties and theorem quoted from Briet, et. al. states that all positive integers map linearly and have the following properties that:

- For any vector $a \in \mathbb{C}^n$, we have $\|f(a)\|_{L_1} \leq \|a\|_{\ell_2}$
- For each standard basis vector $(e_i)$, we have $\|f(e_i)\|_{L_1} = 1$. If $\|f(a)\|_{L_1} > \left(\sqrt{\frac{\pi}{4}} + \varepsilon\right)\|a\|_{\ell_2}$ then $\|a\|_{\ell_4} > (\frac{\varepsilon^2}{K})\|a\|_{\ell_2}$ where $(K < \infty)$ is a universal constant.

With respect to this universal constant, we shall allow ($K$ = 12.511). With respect to these requirements the Hermitian matrices inherently maintain a linear transformation over the real numbers. To conclude the security reduction, with respect to the optimization implications of the Grothendieck inequality extended to non-commutation, semidefinite programming (SDP) has a relaxation of NP-Hard problems related to combinatorial optimization.

The strength of any quantum computer hinges on the advanced capabilities to process large combinations, permutations, and enumerations of sets using superposition of a multi-qubit system. The proposed security reduction of a Grothendieck problem uses this very component of quantum computing to mitigate cryptanalysis based on future and known threats posed by quantum computers by addressing the fundamental computational process as opposed to a specific set of quantum computing algorithms.

Bi-affine, or bilinear matrix inequalities (BMI) allow the use of both vectors and symmetric matrices, opening the path to Pauli matrices based upon a Hermitian metric (Boyd & Vandenberghe, 1997). A linear matrix inequality as a relaxation of a BMI can be defined as the BMI problem introduced by Boyd and Vandenberghe where the BMI problem is expressed as:

$$\text{minimize } c^T x$$

$$\text{subject to } F_0 + \sum_{i=1}^{m} x_i F_i + \sum_{j,k=1}^{m} w_{jk} G_{jk} \geq 0$$

$$w_{jk} = x_j x_k, j, k = 1, \dots, m$$

Within the constraint of($w_{jk}$), if ($j \neq k$) the necessary conditions for bilinear matrix inequalities may be satisfied by quadratic forms. The relaxation of this BMI focuses on the second constraint to derive the linear matrix inequality (LMI):

$$\text{minimize } c^T x$$

$$\text{subject to } F_0 + \sum_{i=1}^{m} x_i F_i + \sum_{j,k=1}^{m} w_{jk} G_{jk} \geq 0$$

$$\begin{bmatrix} W & x \\ x^T & 1 \end{bmatrix} \geq 0$$

A symmetric matrix may be positive semi-definite if the quadratic form associated with the matrix is non-negative. It is useful at this point to refer to the interchangeability of the term eigenvalue with characteristic value, noting for Hermitian matrices that any Hermitian matrix where the eigenvalue is non-negative, then produces a matrix that is positive semi-definite (Cullen, 1990). The semi-definite cone is associated with several semidefinite programming problems, including optimization, denoted as ($S_+$), which is an intersection of half-spaces in the subspace ($S^n$) of the symmetric, semi-definite matrix. Given a symmetric matrix, the square root of the largest eigenvalues which are all non-negative is used to derive the linear operator norm.

The resulting arc length based upon the geodesic is the minimized arc length of the Riemannian metric.

$$\frac{12\pi^6 x^2}{(16x^6 + \pi^6)^{3/2}}$$

*Equation 1 – Maximal Element*

The geodesic depicted results from a center, radius, and equation derived from the holomorphic vector bundle, or the primary function and derivatives of the secondary functions. The center of the geodesic, expressed as:

$$\left(\frac{2x_0(\pi^6 - 8x_0^6)}{3\pi^6}, 12.511 - 0.0752536x_0^4 - \frac{2.58386}{x_0^2}\right)$$

*Equation 2 – Center Point of Geodesic*

$$\frac{(16x_0^6 + \pi^6)^{3/2}}{12\pi^6 x_0^2}$$

*Equation 3 - Radius of Geodesic*

$$(-12.511 + y + \frac{2.58386}{x_0^2} + 0.0752536x_0^4)^2 + \left(x - \frac{2x_0(\pi^6 - 8x_0^6)}{3\pi^6}\right) = \frac{(\pi^6 + 16x_0^6)^3}{144\pi^{12}x_0^4}$$

*Equation 4 – Geodesic Equation*

## Cryptographic Specification

### Design Rationale[1]

Quadratic form allows one to transform a curve by introducing a new coordinate system using a coordinate transformation (Shilov, 1977). A quadratic form is defined on a linear space such that an argument of a vector is obtained by changing $(y)$ to $(x)$ in any bilinear form defined on the linear space. The necessary and sufficient condition for a symmetric matrix in defining positive definite bilinear form requires the descending principal minors of the matrix be positive (Shilov, 1977). In the case of Hermitian matrices of which the bilinear form is symmetric and is based upon a positive-definite form, the bilinear form is derived as a scalar product via substitution, i.e., $\mathbf{B}(x,x)$ is positive definite, the bilinear form $\mathbf{B}(x,y)$ is the scalar product (Shilov, 1977). This results in an orthonormal canonical basis derived from the scalar product, and it is also worth noting that with any non-symmetric bilinear form the derived quadratic form cannot be used to re-construct the bilinear form that generated it (Shilov, 1977).

For any given vector in $\mathcal{L}_3$, the corresponding decision problem of finding a vector which is longer than the shortest vector by a factor of $1 + n^{\varepsilon}$ for the shortest nonzero vector with absolute constant $\varepsilon > 0$, it is NP-Hard if and only if the problem is congruent to the factor constraint of $1 + 2^{-n^4}$ with respect to the lattice with $\mathcal{L}_2$ norm.

Treating a *k*-clique as input, the complexity is generally understood as $2^{\frac{n}{2}}$ for any *n* number of potential *k*-clique sets for $\binom{n}{\frac{n}{2}}$, which results in the input length $|x| + |k| = n + \log n$ for

Extending the verification of key strength, the use of the **closest vector problem** (CVP) and the **shortest vector problem** (SVP) introduce further complexity in conjunction with both greater resilience in implementation as well as higher adaptability in terms of adjustments to the overall scheme.

The SVP states, for a randomized reduction with constraints of a lattice $\mathfrak{L}$, such that the norm of the lattice is 1.82 with respect to the constant *e*, where $e > O$, finding a vector of length longer than the shortest vector by a factor of $1 + 2 - n'$ is NP-Hard, and NP-Complete for randomized reductions in terms of decidability (Ajtai, 1998).

Osculating circles occur at a specific point, the uniqueness of which allows noise within the polar coordinates of the holomorphic vector bundle. The use of sufficient minute values of noise within lattice-based cryptography confirm that it is an acceptable post-quantum cryptographic scheme.

To counter the failings of Lattice-Based Cryptography (LBC), the key space must be sufficient to withstand quantum computing threats, but also feasibly be implementable for wide-spread production and use. If a linear manifold is finite, the search space for noise as perturbation within a finite vector space produced by the inner product of the manifold is reduced. Furthermore, the inner product of two vectors may be a tensor that is commutative (Budge, 1991). Regarding the implementation of noise, as well as the level of noise, the smaller search space may enable post-quantum capable encryption, as well implementation in large-scale networks.

Per Shilov in his book *Linear Algebra*, on page 41 section 2.35.d he states that "…every complex linear space…is twice as large as a complex linear space if it is regarded as a real instead of complex" (Shilov, 1977). Therefore, whether the decision is to use a larger space by regarding it as a real-valued system rather than complex, or vice versa the ability to influence the overall key space for optimal security and performance is straightforward. If a larger space is needed once it could be discovered that the key space is insufficiently sized, a switch between the systems of ($\mathfrak{R}$) or ($\mathfrak{C}$) could hypothetically solve this problem.

**Design Decisions**

The proposed advances to the TLSv1.2 network protocol offer the following improvements within the context of a post-quantum resilient scheme. Through use of three NP-Hard problems, adjusted as NP-Complete in terms of randomized search decidability, the goal is stated as mitigating hybrid attacks from implementations of Shor's factoring algorithm, and Grover search algorithms for known and unknown numbers of solutions. The given problem is as follows. For any arbitrary bilinear quadratic in $\mathbb{C}^4$, find any set of points in $\mathbb{C}$ such that the original graph of the bit map is calculable using only the homeomorphism of any arbitrary $k$-clique of unknown network $(G)$.

The proposed scheme reduces the $k$-clique problem as an input function, resulting in the generation of a quadratic used as parameters within the lattice. By virtue of Grothendieck's inequality, which is the principle NP-Hard problem for the security reduction, the satisfiability of lattice constraints in terms of NP-Hard and NP-Complete bounds is provably congruent to a closest vector problem in lattice $\mathcal{L}_3$, where the base vectors of $\mathcal{L}_3$ are treated as a holomorphic vector bundle, denoted as the unit ball with congruent topology to the Riemann sphere.

For the Grothendieck constraints, the relative vector norms of $|x| < 1$, $|y| < 1$ necessarily results in satisfaction of NP-Hard requirements for shortest vector problems in the lattice $L_2$. While research in Ring Learning with Errors (R-LWE) schemes have pushed to produce a lattice structure from a single, small vector as both secure and efficient are widely seen as a failure in security, the scheme proposed in this verification hinges on the use of vector bundles such that a principle bundle may be derived. This principle bundle, treated as the unit sphere has as homeomorphisms, an equivalence of the derivatives shared by two holomorphic vectors, wherein the remaining holomorphic vector may be trivially derived.

In this manner, the achievement is such that a single vector bundle may be applied as the base vector, while simultaneously capable of being folded into any of the other two vector bundles. The resulting coordinate shift achieved through a tensor produces stable security in addition to optimal costs in terms of use.

The entire set as the complete system shall be defined as open, resulting in an empty and open complement which therefore defines the system as closed and open. The primary analytic function is taken from Malloy (Malloy, 2016), and will be applied throughout the methodology as the principle expression adapted to the maximal element. It was determined in Malloy that the existence of this analytic function is unique, or holomorphic, and a continuation of an Abelian-Banach space within a Riemann-Hilbert intersection (Malloy, 2016). These properties will be assumed throughout the methods of the security reduction described herein. Given that the system is within a Hilbert space, the Grothendieck constant is inherent within the scope of this novel cryptographic scheme. This has as a requirement, which needs to be satisfied, norms relative to the vectors of $(x_1, x_2, \ldots x_m)$ and $(y_1, y_2, \ldots y_n)$ that $|x_i| \leq 1$ and $|y_j| \leq 1$.

For the purposes of this reduction, the vectors $(x_i y_j)$ with respect to covariance maintain symmetry. Therefore, any adjustment to either vector which removes the co-variance may result in an inequality derived from the resulting asymmetry which would also introduce an inability to reconstruct the generative bilinear form based on any quadratic form which may follow. This capability is achieved by treating $\varepsilon$ as a permutation tensor as well as an introduction of arbitrarily small noise. Such use of $\varepsilon$ results from treating $(\theta)$ as a point of reflection or curl based upon polar coordinates.

The manifold $(M)$ is a Riemannian manifold boundedly compact. Given this, $(M)$ possesses a metric tensor, where the metric shall be defined per its inner product of the tangent space. Within the tangent space, the Riemannian metric results, thereby introducing the positive-definite metric tensor along with the real-valued metric. Given any metric space $(R)$ within $(M)$, conditions for boundedness are for an $\varepsilon$-net, such that $(x \in M)$ and $(a \in A)$ where$(A \subset R)$, there exists a point $p(a, x) < \varepsilon$ (Kolmogorov & Fomin, 1999). The value $p(x)$ may be selected arbitrarily if there exists at least one $p(a)$ which satisfies this condition (Kolmogorov & Fomin, 1999). Boundedness is both necessary and sufficient for compactness (Kolmogorov & Fomin, 1999). With respect to epsilon, it will act as both permutation tensor and arbitrarily small values of noise. The value of $p(a)$ shall be the arc length derived from the geodesic, ensuring satisfaction of the constraint that $p(a, x) < \varepsilon$ given that the upper bound of the vector is$|x_i| \leq 1$ and $(\varepsilon \leq 2)$.

The Riemannian metric may be regarded as weak, implying a positive definite property while allowing looser requirements for Riemannian metric satisfiability such as isomorphism between the tangent and cotangent spaces. The geodesic of $(M)$ is graphically represented as osculating circles to calculate the shortest arc length of the maximal element to derive the tangent bundle. The analytic function, Equation 1, introduces the principle expression for the security reduction.

$$\left(12.511 - \frac{z^4}{\pi^3}\right)$$

*Equation 5 – Principle Holomorphic Expression*

The canonical Hermitian inner product so defined by Equation 1 is a result of the matrix $(E)$ defined as the identity matrix. Equation 1, as was determined by Malloy, produces a symplectic form on $(M)$ over a holomorphic vector space by the complex Hilbert space requirements satisfied by Malloy to derive the analytic function (Malloy, 2016). The necessary conditions of the Hermitian inner product specific to reduction of the Grothendieck problem chosen requires commutative properties of the inner product for relaxation of the Bilinear Matrix Inequality problem (BMI) (Boyd & Vandenberghe, 1997). The BMI shall be implemented to reduce Grothendieck's Non-Commutative (GNC) problem for the proposed cryptographic scheme.

The Hermitian inner product generically defined has a symmetric positive definite real part, while the imaginary component is symplectic. The avoidance of defining $(E)$ as a Hilbert identity was chosen to provide a unitary basis more easily, as well as introduce the Pauli matrices. By the tangent bundle, a Jacobian may be derived as an iterative method for decimal floating point operations. This allows use of $(\mathbb{Q})$ to reduce the integer $(2)$ accordingly, relative to the constant$(k_R(n))$ and the vectors$(x_m, y_n)$.

The appropriate $(n)$ value with respect to $(2)$ within the context of the GNC problem is traditionally known to be$\left(\sqrt{2}\right)$, and any value lesser than $(2)$ results in NP-Hard approximations (Briet, Regev, & Saket, 2015). In a proven refutation of Grothendieck's conjecture for complex matrix $(A)$ of which the numbers$(s_i, t_j \in A)$ for constant $\left(k_C(n)\right)$ when $(n = 2)$ are known to satisfy $[1.1526, 1.2157]$ (Finch, Mathematical Constants). The two values which satisfy the constant $(k_C(n))$ may be substituted within the vectors$(x_m, y_n)$.

Equation 2 is the image of $f(H)$. Figure 1 is the graphical representation of Equation 2. Within the parametric plot range between$(0, \pi)$, for the domain of $(0, 2)$ the initial entropy derived from the modified analytic function produces the image for projection.

$$\left(12.511 - \frac{n^4 x}{\pi^3}\right) = (0, \pi, \infty)$$

*Equation 6 – Image*

Analyzing Equation 2 as a preimage of $f(H)$ results in intersections of all roots in the complex plane equivalent to all roots of Equation 1. The resulting intersections allow implementation of surjective functions within the roots of both Equation 1 and 2, from the maximal element. The arc length within the range of $(0, 2\pi)$ relative to the parametric curve, with respect to the doubling of the roots of Equation 1 approximately equal the value of the parametric arc. The parametric curve is generated from an adjusted Gaussian approximation to prime numbers, denoted as Equation 3 in complex form. From the doubling of the roots of Equation 1, the approximation to the value of the arc allows the introduction of torsion with respect to parametrization such that one may choose either the vector of the geodesic or instead operate as if a particle is accelerated.

For the purposes of this reduction, the polar graphs produced by the holomorphic vector bundle shall be termed "onions." Given the basis of each onion on a Bloch sphere, the introduction of acceleration to a point within the geodesic generates a shift in coordinate systems. If a particle, or point, is treated analogous to a Bloch sphere then within each point the respective position determines specific behavior. Bloch spheres have as a property that a pole may either act to generate or destroy a particle. Therefore, any point within the geodesic shares this property such that once acceleration is introduced, the coordinate system changes from an initial value to one based upon the vector, trajectory, and initial vectors. This is to introduce asymmetry to generate an inequality within bilinear matrices as a method to obfuscate the base similar to the closest vector problem.

To achieve the ability of applying arc lengths properly, a parametrized function must be introduced. The proposed parametric function shall be based upon a Gaussian expression calculated from Gauss' approximation to twin primes, as defined by Malloy (Malloy, 2016). This is shown as Equation 3 in complex form.

$$ie^{-i\theta} - ie^{i\theta}$$

*Equation 7 – Gaussian Expression*

The Gaussian expression permits introduction and use of arbitrarily sized integers as a co-variant of the period $(2\pi)$ for the integer roots$(\theta: 0, \pi)$. This transcendental function can be expressed trigonometrically as a function of $sin$ and $cosin$ in another representation, using the parameters of $(0, 2\pi)$ as a range for the variable$(t)$.

$$2\sin(t), \qquad \begin{Bmatrix} \sin(t)\ ) \\ \cos(t) \end{Bmatrix}, \qquad t = 0\ to\ 2\pi$$

*Equation 8 –Parametrized Gaussian Expression*

The initial arc length of $(M)$ is derived from parametrization of the Gaussian expression, which shall be demonstrated not to be the smallest arc length of$(M)$.

$$\int_0^{2\pi} \sqrt{4\cos^2(t) + \sin^2(t)}\, dt = 8E(\frac{3}{4}) \approx 9.68845$$

*Equation 9* – Gaussian *Arc Length*

$$ie^{179.21°(-i)} - ie^{179.21°(i)}$$

*Equation 10* – Gaussian *Polar Coordinate*

The polar graphs shown in Figures 5-8 for ranges of $(z)$ based upon Equation 1 result from an increase of the respective range of$(z)$, listed as the set$(A)$.

$$A = \begin{Bmatrix} (-\pi, \pi) & (-20{,}20) \\ (-37, 37) & (-100{,}100) \end{Bmatrix}$$

*Equation 11- Holomorphic Function Range*

Each range of $(z)$ produces a series of expanding arcs from the point of origin at approximately 0, and as shown by the graphs of the principle analytic function, a slight perturbation in the ball results from an increase in range, slightly distorting the ball.

To revisit the parametric curve, the given parameters for the algorithmic expression denoted as the adjusted principle holomorphic function intersects at the signed values of $(-1, 1)$ upon the *y-axis*. The use of this intersection upon the *y-axis* is to allow for introduction of slight noise as "pulses" of the holomorphic vector bundle. The concept of "pulse" shall be regarded as a layering effect of the encryption upon the cipher text based upon values of additive noise derived from $\varepsilon$-permutations.

The holomorphic vector bundle is a set of functions where each function produces a unique orientation of its respective polar graph. The uniqueness in orientation, as well as the homeomorphism between each, lends itself to problems such as the closest vector problem.

$$\begin{cases} f(x, y, z) = \left(-\dfrac{4z^3}{\pi^3}\right) \\ g(x, y, z) = \left(12.511 - \dfrac{z^4}{\pi^3}\right) \\ h(x, y, z) = \left(-\dfrac{z^4}{\pi^3}\right) \end{cases}$$

*Equation 12 – Holomorphic Vector Bundle*

Between each onion, the primary difference is orientation, the other being the range of each function. Apart from differences in range and orientation, the domain remains the same with respect to all three holomorphic functions. The applications of each onion is presented as an ability to structure bilinear forms within and between functions and composite functions such that identification of the bilinear form based on the generated quadratic form is difficult. The extension this has towards the security reduction applies the bilinear form as a matrix inequality such that the roots of each function is applied as a value of $|x_i| \leq 1$ and $|y_j| \leq 1$.

Further applications of the quadratic form extend to changes in coordinate systems between each polar coordinate derived from the holomorphic vector bundle. By projecting the concentric circles through the hyper-plane, an osculating circle for $(x_0)$ enables the calculation of both the maximal element as well as reduction of the parametrized arc length based upon the geodesic. The resulting arc length based upon the geodesic is the minimized arc length of the Riemannian metric.

Given the **bilinear matrix inequality** (BMI) constraint of the relative norm of the *y* vector, and the same constraint applied to the relative norm of the *x* vector, this condition for the BMI is satisfied with respect to the *y* vector for positive values of *n*. Insofar as satisfying the same condition for the values of *x* with respect to $n \in \mathbb{Z}$, the primary number base for *x* is zero. Otherwise, the parent function of **X** for $n \in \mathbb{Z}$ is expressed as:

$$\mathbf{X}: e^{-ix} - 2i = -2i + \sum_{k=0}^{\infty} \frac{(-ix)^k}{k!}$$

With *y*-intercepts located at:

$$(0,1) \in \mathbb{R}, (0,-2) \in \mathbb{C}$$

Given the root of **X**:

$$\frac{1}{2}(4\pi n - \pi + 2i \log 2)$$

The utility of the series of **Y** as a period of 2 with respect to the *y*-intercepts of **X** allow efficient use of the TAP algorithm. The TAP algorithm couples $(x, y, \theta)$ between the range of values for $y$ along a dependent axis, where the dependent axis is a transition vector used by TAP. Since TAP occurs with respect to all zeroes of the system, TAP results in a coordinate shift between **X** and **Y** for $(\mathcal{O}, \mathfrak{T})$. The series **X** is periodic in $(x)$ with period $2\pi$ while the series **Y** is periodic in $(y)$ with period 2. Both series are with respect to a universal constant, $K = \infty$.

The series representation of **Y** is equated to the following function:

$$\mathbf{Y}: 2e^{-y\pi i} = 2\sum_{k=0}^{\infty} \frac{\pi^k(-iy)^k}{k!}$$

No roots exist for the function **Y**. The reduction of the non-commutative Grothendieck inequality (NCG) centers on vector constraints and conditional values of tensor products applicable through the scheme's architecture. For a given linear operator $\mathcal{F}$, and the vectors $w_{j,k}$, **X**, **Y**, the following holds. For any arbitrary vector $X = \begin{bmatrix} x_i \\ x_j \\ x_m \end{bmatrix}$, the tensor product of $X \otimes Y$ as permutation of the vector $Y = \begin{bmatrix} y_j \\ y_n \\ y_m \end{bmatrix}$ may be expressed as

$$X \otimes Y = \begin{cases} x_j, y_j \text{ for } j \neq k \\ w_{jk}, \text{otherwise} \end{cases}$$

Insofar as the vector of $w_{j,k}$ has the additional constraints of:

$$w_{j,k} = \begin{cases} x_{j,k} \\ j \\ k \end{cases}$$

The given linear operator is a function of the vector of $w_{j,k}$, treated as a bijective function:

$$w_{j,k} \xrightarrow{\mathcal{F}} x_0 = \begin{bmatrix} x_j \\ y_j \end{bmatrix}$$

For any constant, $k = 1, \ldots, m$ when $K = \infty$, such that $n = (-1,1)$, $x = 0$, and $p: (x = n) \rightarrow (n = 0)$.

Therefore, this geodesic point is a quadratic zero periodic in $Y$, with bilinear base $(X, Y)$. Topological properties place the geodesic in the neighborhoods of $(x, y: 2, \text{-}\,2)$. Inverting these values of $x, y$ and $(2, -2)$ such that $(x = -2)$ and $(y = 2)$ is a TAP for the Blue onion in this RACK. The TOP is pictured as the intersections of the points $(B, C)$. The relative pole orientation generates the polarity as a vector bundle denoted as $\vec{\alpha}$, signifying that this bundle produces values analogous to an alpha polarity of a Bloch sphere. A $\vec{\beta}$ polarity is the opposing, cancellation vector bundle, where no vector tends towards these values, yet marks the initial point of transition towards $\vec{\alpha}$.

**Algorithm Collection**

Bob wishes to share an encoded message with Alice. This message is encrypted with Alice's public key, and only Alice's private key can decode any message sent using her public key. Alice's private key is a quadratic form of arbitrarily chosen onions, $(O_{A,B,C} \in \mathcal{O})$ called an "O-Clique," and represented as a graph in matrix form. To produce the O-Clique, the onions are converted into a matrix, and then translated into the secret $(s)$ which is a quadratic form. Once this is complete, the clique is "PEN'd" to add arbitrary nodes within the graph. The H-Clique is then converted to polynomial form and obfuscated by shifting weights. The resulting clique is now Alice's TOP expressed as $\text{TOP}_A$. Using $\text{TOP}_A$, Alice then shares her public key, denoted as BLIP. Bob then uses BLIP to encode his message to Alice. Alice takes the message from Bob and extracts the message by using TAP operating upon BLIP with respect to TOP. Since TOP is a vector field, using TAP with BLIP produces the H-Clique, the final clear text must be produced by an ephemeral key Alice generated by her choice of onions for the O-Clique. The quadratic form from the onion matrices used to create the O-Clique is incalculable from the quadratic. Since Alice is the only one who knows the quadratic form and base matrices, only she knows which nodes to subtract from the H-Clique. The following diagram describes this process:

Protocol Algorithm Terms

- TEQ – Tensor Edge Quadratic
    - Alice Handshake
    - Key Exchange Pulse
- MIQ – Matrix Inequality Quadratic
    - Bob Handshake
    - Key Exchange Pulse
- NET – Node-Edge Tensor
    - Handshake Negotiation
    - Trapdoor
- PEN – Point Edge Nodes
    - Generates Arbitrary Nodes upon Edges

General Asymmetric Terms

- TOP – Tensor Origin Point
    - Bilinear Vector Field from Intersecting Onions used by TAP
- BLIP – Bilinear Intersection
    - Public Key
- TAP – Tensor Access Point
    - Linear Operator $\otimes$
    - Generates Nodes for TOP
    - Creates Homeomorphism of O-Clique
- $\varepsilon$-net filter
    - Permutation Tensor Added to TAP

To describe the process diagram, the following steps are the operations needed to implement the asymmetric onion scheme.

1: Alice Chooses Onions, Cliques – “Alice Racks the Onions”

2: Alice encrypts her key by generating a homeomorphism of the rack – “Alice PENs her Key”

3: Taking the H-Clique, Alice then uses TAP to create TOP – “Alices TAPs the Clique”

4: After the TOP vector field is produced, Alice can transmit her BLIP – “Alice’s BLIP is up.”

5: Bob uses Alice’s BLIP to encode his message – “Bob BLIPS Alice”

6: Alice Authenticates the BLIP using TOP – “The BLIP TOPs”

7: Alice pairs the TAP, TOP, and BLIP to get the H-Clique – “Alice TAPs the BLIP”

8: Alice uses the H-Clique and PEN to decode using her O-Clique – “Alice TAPs the PEN”

9: Alice now has the BLIP message from Bob.

## Algorithm One – Tensor Origin Point

### TOP Parameters

Given the use of tensors, by default they are regarded as a vector field resulting from the intersection of vectors. The characteristic of a tensor as a vector is derived from the membership between a dynamic, interacting system. With each onion based upon a particle, constructed as a Riemann sphere which itself is modeled in quantum computing as a Bloch sphere, the respective poles of an onion have specific properties associated with them.

### Mathematical Operations and Equations

Onion A ($O_A$) + Onion B ($O_B$) using TOP is outlined as follows:
$O_A$ is an N x N matrix
$O_B$ is an M x M matrix
TOP is the incident N x M matrix

1 - $O_A$ PEN $O_B$ creates nodes for intersection
2 - $O_A$ TAP $O_B$ and $O_B$ TAP $O_A$ begin generating edges between nodes
3 - The vector fields produced by step 2 become TOP

Given the use of tensors, by default they are regarded as a vector field resulting from the intersection of vectors. The characteristic of a tensor as a vector is derived from the membership between a dynamic, interacting system. With each onion based upon a particle, constructed as a Riemann sphere which itself is modeled in quantum computing as a bloch sphere, the respective poles of an onion have specific properties associated with them. The following is the TAP algorithm:

$$\left(\vec{a} \in \mathrm{TOP}_A\right) \text{ and } \left(\vec{b} \in \mathrm{TOP}_B\right)$$

$$\left(\vec{a} \otimes \vec{b}\right) \in \left(\mathrm{TOP}_A \otimes \mathrm{TOP}_B\right)$$

$$\mathrm{TOP}_A \otimes \mathrm{TOP}_B \rightarrow \overrightarrow{\mathrm{TOP}}_{A,B} = \mathrm{TOP}_{A,B}$$

$$\mathrm{TOP}: (\mathcal{O}) = \begin{cases} \mathrm{BLIP}_{N,M} \\ x_0 \in O_B \\ \mathrm{TAP}: \theta, y \ \in \mathcal{O} \end{cases}$$

**Algorithm Two – Bilinear Intersection Point (BLIP)**

**BLIP**

Even though $O_A$ and $O_C$ only have $(0)$ as a root, $O_A$ has an odd parity while $O_C$ has an even parity. The parity of $O_B$ is even, and both $O_B$ and $O_C$ have a global maximum of 0 at $z = 0$. Thus, a BLIP is achieved by using the parent functions of the sets $\{X\}$ and $\{Y\}$, denoted as **X** and **Y**, but remains secure when released publicly**.** Since the onion choice and derivative chosen is unknown, the parity of the key pair is unknown. Since the parity of the key pair is unknown, having both BLIP and TOP will not reveal which TAP is needed. Even if a TAP is found in the TOP vector field, since the H-Clique was PEN'd and had weights adjusted, the actual O-Clique remains unknown. Given the O-Clique is a quadratic form used to generate the bilinear matrices, knowing the matrices will not permit calculating the quadratic O-Clique either. By using this method, the key selected as the quadratic form is ephemeral.

### Mathematical Operations and Equations

The BLIP algorithm relies on the TAP operator, ⊗, which manipulates the zeroes of the set $O$ by taking the first derivatives of $(O_B, O_C)$ such that:

$$\left.\begin{matrix} O_B{}' \\ O_C{}' \end{matrix}\right\} = O_A$$

The principle expression for $O_A$ is:

$$O_A : -\frac{4z^3}{\pi^3}$$

While the first derivatives of $O_{B,C}$ are equal, the onions $O_{A,C}$ share the same root of (0), this being the only root of the onions $O_A$ and $O_C$.

## Algorithm Three – Tensor Origin Point (TOP)

### Mathematical Operations and Equations

## Algorithm Four – Tensor Access Point (TAP)

### TAP

The TAP function produces an intersection between real and imaginary values, where the real intercepts occur at (-1, 1) as a value for $(n)$, where $n = x$, and exists along the *x-axis.* The imaginary values of TAP occur at $(-1, -4)$ and $(1, -4)$ again using the equation $(n = x)$. The convergence of the real and imaginary values of TAP occurs at the point of origin, (0).

Given the BMI constraint of the relative norm of the *y* vector, and the same constraint applied to the relative norm of the *x* vector, this condition for the BMI is satisfied with respect to the *y* vector for positive values of *n.* Insofar as satisfying the same condition for the values of *x* with respect to $n \in \mathbb{Z}$, the primary number base for *x* is zero. Otherwise, the parent function of **X** for $n \in \mathbb{Z}$ is expressed as:

$$\mathbf{X}: e^{-ix} - 2i = -2i + \sum_{k=0}^{\infty} \frac{(-ix)^k}{k!}$$

With *y*-intercepts located at:

$$(0,1) \in \mathbb{R}, (0,-2) \in \mathbb{C}$$

Given the root of **X**:

$$\frac{1}{2}(4\pi n - \pi + 2i \log 2)$$

The utility of the series of **Y** as a period of 2 with respect to the *y*-intercepts of **X** allow efficient use of the TAP algorithm. TAP couples $(x, y, \theta)$ between the range of values for *y* along a dependent axis, where the dependent axis is a transition vector used by TAP. Since TAP occurs with respect to all zeroes of the system, TAP results in a coordinate shift between **X** and **Y** for $(\mathcal{O}, \mathfrak{T})$. The series **X** is periodic in $(x)$ with period $2\pi$ while the series **Y** is periodic in $(y)$ with period 2. Both series are with respect to a universal constant, $K = \infty$.

The series representation of **Y** is equated to the following function:

$$\mathbf{Y}: 2e^{-y\pi i} = 2\sum_{k=0}^{\infty} \frac{\pi^k(-iy)^k}{k!}$$

No roots exist for the function **Y**.

**Mathematical Operations and Equations**

The TAP function is expressed as:

$$2e^{-i\pi n} - 2e^{i\pi n}$$

The following is the TAP algorithm:

$$\left(\vec{a} \in \text{TOP}_A\right) \text{ and } \left(\vec{b} \in \text{TOP}_B\right)$$

$$\left(\vec{a} \otimes \vec{b}\right) \in \left(\text{TOP}_A \otimes \text{TOP}_B\right)$$

$$\text{TOP}_A \otimes \text{TOP}_B \rightarrow \overrightarrow{\text{TOP}}_{A,B} = \text{TOP}_{A,B}$$

$$\text{TOP}: (\mathcal{O}) = \begin{cases} \text{BLIP}_{N,M} \\ x_0 \in O_B \\ \text{TAP}: \theta, y \ \in \mathcal{O} \end{cases}$$

Example of Tapping:

$$\text{RACK}: \begin{cases} O_A \\ O_B \end{cases}$$

$$\text{PEN}: = G(O_A) \oplus (O_B)$$

$$\text{TAP}_{A,B}: \begin{cases} O_A \\ O_B \end{cases}$$

$$\text{TAP}_A = O_A: (x, n)$$

$$w_{j,k} = \begin{cases} x_{j,k} \\ j \\ k \end{cases}$$

$$w_{j,k} \overset{\mathcal{F}}{\rightarrow} x_0 = \begin{bmatrix} x_j \\ y_j \end{bmatrix}$$

$$k = 1, \dots, m$$

$$n = (-1,1) \text{ and } x = 0$$

$$p: (x = n) \rightarrow (n = 0)$$

$$\text{TAP}(O_A, O_B) \bigotimes \text{TOP}\ (\mathbf{X}, \mathbf{Y})$$

$$X = \begin{bmatrix} x_i \\ x_j \\ x_m \end{bmatrix} \ Y = \begin{bmatrix} y_j \\ y_n \\ y_m \end{bmatrix}$$

$$X \otimes Y = \text{BLIP} \begin{cases} x_j, y_j \text{ for } j \neq k \\ w_{jk}, \text{otherwise} \end{cases}$$

It has been established that $(O'_B, O'_C) = O_A$, and that for any *x* with a base of 0, the relative Onion is $O_B$. The PEN algorithm takes any sub-graph *k*-clique as the O-Clique, and proceeds to introduce arbitrary nodes upon arbitrary edges to generate the homeomorphic clique of the O-Clique. This new homeomorphic clique, termed H-Clique is then TAP'd to eventually become a public key. The O-Clique itself is an ephemeral key, obfuscated by PEN and resulting in the encrypted H-Clique which is then diffused in the TOP by a TAP.

The process of encrypting your key relies on the parametrization of the system $\mathcal{O}$, which is achieved with the function:

$$ie^{-i\theta} - ie^{i\theta}$$

This step is referred to as "Penning your Key." The utility of this function is based upon the periodicity, which is a period of $2\pi$ being periodic in $\theta$, with the integer roots of $\theta$ being 0 and $\pi$. The roots of this function are $\pi n$, otherwise. Therefore, PEN operates by beginning at the zero of the system, which is the center of the geodesic based in $O_B$.

**Algorithm Five – Point-Edge Node (PEN)**

**PEN Algorithm:**

The start state begins in base zero for the PEN algorithm, applied as a function to the O-Clique. This begins the process to manipulate the chosen ephemeral key regardless of the ephemeral key selected. The ability to transform any O-Clique into an H-Clique is achieved via the center point of the geodesic, derived from $O_B$, along with the ability to shift coordinate systems between any onion by simply either taking the derivative of the parent onion or doing nothing. With the start state in base zero, the $x$-coordinate then becomes the value of the geodesic center point. With this set, the relative $y$-intercept at $(x = 0)$ can then either be treated as a complex or real number. Given that the period of 2, periodic in $y$ is the relative properties to the value of $x$ at $\mathrm{PEN}_0$, a PEN-Step is therefore defined as an iteration of one period of 2. Since $x$ as defined by the parent function **X** is periodic in $x$ with period $2\pi$ it is straightforward to operate within both parent functions $(\mathbf{X}, \mathbf{Y})$ with respect to the parametric function. The next stage begins upon completion of the step-cycle permutations, and is a result of an additive union of the positive integers to the O-Clique. This unity stage between the integers and the O-Clique is calculated using the base $(0, \mathbf{Y})$ for PEN, symbolized as $\mathrm{PEN}_{0,\mathbf{Y}}$. The final stage of the PEN algorithm results once the "PEN is Tapped," meaning a "Tensor Access Point" is added to the O-Clique, producing the newly formed homeomorphic H-Clique.

**Mathematical Operations and Equations**

***PEN Start State:***

$$\mathrm{PEN}_0 = \frac{2x_0(\pi^6 - 8x_0^6)}{3\pi^6}$$

Penning the Key:

$$\mathrm{G}(\mathcal{O})\begin{cases}\mathrm{PEN}_A\\ \mathrm{PEN}'_B\\ \mathrm{PEN}'_C\end{cases}$$

PEN Steps:

$$\mathrm{PEN}_{0,\mathbf{Y}}\colon \mathbb{Z}^{+} \uplus \mathrm{G}(\mathcal{O})$$

Tapping the PEN:

$$\mathrm{TAP} \overset{\otimes}{\rightarrow} \mathrm{PEN} = (\mathrm{H})\mathrm{Clique}$$

***PEN Computations:***

$$\mathrm{PEN} \coloneqq \mathrm{Step}_{\mathbf{Y}} \overset{\$}{\rightarrow} \begin{cases}1, \text{place node}\\ 0, \text{do nothing}\end{cases}$$

The PEN continues the PEN steps and computations for as many iterations, or periods, desired. Essentially the PEN begins as a point in the center of an onion, and then proceeds to "walk" the circumference of the parametrized system $\mathcal{O}$, and based upon the conditions of the relative topology to the O-Clique, in addition to $\mathfrak{T}$ a node is either added between the edge, resulting in the XOR of that position in both the adjacency matrix and incidence matrix, or nothing is done. Once the PEN has been "tapped," the output becomes a *z*-coordinate that uses an XOR of the H-Clique based upon the coordinates and equation for the geodesic.

**Tunable Parameters**

1 - Rack Size

2 - Color of Onions

3 - Integer Layering

4 - PEN Steps

## Performance Analysis

## Known Answer Test (KAT) Values

## Expected Security Strength

For any arbitrary homeomorphism, there exists a holomorphic vector bundle as a range within a parametrized Gaussian domain in $\mathbb{C}$. Knowing any solution to the problem is necessarily within these constraints, the worst-case time complexity may only be derived by first isolating the closest vector, and then calculating the shortest vector for the base bilinear $\binom{0}{1}$-Matrix. We show that the least lower bound (LLB) in queries to an arbitrary Grover oracle requires $O(\sqrt{N})$ evaluations for finding solution $k_0$ (Pittenger, 2000). To increase the probability of success in queries $t$ for $t$ solutions, the work factor of a function $f$ to return a decision is $O(\sqrt{N/t})$ (Pittenger, 2000).

## Algorithm Resistance to Known Attack Vectors

To mitigate attempts of cryptanalysis using Shor's algorithm, the TAP algorithm is populated with high values of noise. The following figure depicts where the levels of noise begin, as well as their respective number field. For hybrid attacks using Grover's algorithm, assuming the encoded search space for Grover has been attenuated for noise, the intractability of solving for the correct derivative, and therefore racked onions remains successful. The derivatives applied to the O-Clique do not provide the quadratic basis for the BMI, and therefore any implementation of Grover will not know which domain to search even with the range known. Given that the vector fields of the TOP and BLIP rely on a TAP, if one were to isolate the TAP from the TAP function they would still be unable to apply the correct PEN an isolate the needed O-Clique. With the added fact that each PEN is ephemeral, there is no risk posed to reusing the PEN.